# Time-Resolved Spectroscopy of Single Excitons Bound to Pairs of Te Isoelectronic Impurity Centers in ZnSe

A. Muller,[1] P. Bianucci,[1] C. Piermarocchi,[2] M. Fornari,[3] I. C. Robin,[4] R. André,[4] and C. K. Shih[1*]

[1]*Department of Physics, University of Texas at Austin, Austin, Texas 78712*
[2]*Department of Physics and Astronomy, Michigan State University, East Lansing, Michigan 48824*
[3]*Department of Physics, Central Michigan University, Mt. Pleasant, Michigan 48859*
[4]*Laboratoire de Spectrométrie Physique, Université J. Fourier, 38402 St. Martin d'Hères, France*

(Dated: March 16, 2005)

Tellurium impurity centers in ZnSe were individually probed with time-resolved photoluminescence (PL) spectroscopy. Resolution-limited peaks with an ultra-low spatial density originate in the recombination of excitons deeply bound to isolated nearest-neighbor isoelectronic Te pairs (Te$_2$). This interpretation is confirmed by ab-initio calculations. The peaks reveal anti-bunched photon emission and a doublet structure polarized along [110] and $[\bar{1}10]$. We analyze the time-resolved PL decay to clarify the role of the dark states in the spin relaxation and radiative recombination of single fine-structure split excitons.

The fundamental physical properties [1] and device applications [2] of three-dimensionally confined solid-state systems have generated tremendous research efforts in the last decade. The quantum optical properties of single emitters have received particular consideration [3], and drawn interest from both atomic and condensed matter physics. Emphasis has been placed on semiconductor quantum dot (SQD) systems, in which excitons are confined to mesoscopic length scales and closely mimic properties previously unique to isolated atoms or molecules.

Lately, single impurity states in bulk semiconductors have attracted renewed attention for quantum optical studies. It has been argued that using impurity states, most technological advantages of solid-state "artificial atoms" are retained while avoiding difficult material issues facing the SQD community. Indeed, recent demonstrations of non-classical light emission from individual N-vacancies in diamond [4] and single N atoms in ZnSe [5], show that impurity states are good quantum emitters. Moreover, optical spectroscopy of single N pairs in GaAs has been reported [6], exposing features unique to *homogeneously* broadened emitters, such as their configuration, polarization anisotropy, and spatial distribution.

Here we demonstrate time-resolved PL spectroscopy of individual Te impurity pairs in ZnSe (Te$_2$). Te$_2$-bound excitons combine three characteristics that make them desirable as single emitters when compared to the systems above. (i) The confinement of the exciton is in the range 40-100 meV, providing a large spectral separation from the free exciton emission. (ii) Isoelectronic impurities are similar to neutral epitaxial SQDs since they do not permit Auger exciton recombination. The latter is known to dominate for excitons bound to acceptors and donors [7], and should also strongly affect the recombination of trions in charged dots. (iii) Since the excitons are strongly bound to Te$_2$ but not to Te, the density of single emitters is reduced drastically to less than 5 μm$^{-2}$. This eliminates the need of submicron apertures or mesas, and makes the system appealing for integrated nano-photonics where single photon emitters need to be coupled to optical nanocavities and waveguides.

The absence of inhomogeneous broadening permits unprecedented insight into this system. A biexponential PL decay reveals information about the exciton fine-structure that was previously completely concealed in ensemble measurements. In addition, photon anti-bunching provides unequivocal proof that the sharp emission indeed originates from single quantum entities, which was suggested, but not proven, in the case of GaAs:N$_2$ [6]. Together with the energy structure readily visible in the polarized PL spectrum, a unified picture of the excitonic fine-structure is available.

The sample investigated was grown by molecular beam epitaxy (MBE) on a GaAs substrate and consists of 1 monolayer (ML) Te-doped ZnSe sandwiched between 40 nm thick layers of ZnSe, with a nominal Te concentration of 2500 μm$^{-2}$. Tellurium is known to isoelectronically substitute Se, providing unusually strong confinement [8, 9]. The sample, typically maintained at 10 K, was excited non-resonantly (415 nm) with a frequency-doubled Ti:Sapphire laser that delivers ~ 1 ps pulses at a repetition rate of 80 Mhz. The PL was dispersed by a 0.5 m spectrometer and imaged onto a two-dimensional charge-coupled device (CCD) detector [10]. Photon correlation measurements were performed using a conventional Hanbury Brown and Twiss (HBT) setup, with a 50/50 beam splitter and two single-photon-counting avalanche photodiodes (APDs) [3]. Such a setup measures the number of photon pairs $n(\tau)$ with arrival time separation $\tau$, which is proportional to the second order correlation function $g^{(2)}(\tau)$. For time-resolved single photon measurements, the same setup was used with only one APD (time resolution ~500 ps). When higher collection efficiency was necessary, we used a high-index hemispherical solid immersion lens [11] in direct contact with the sample surface.

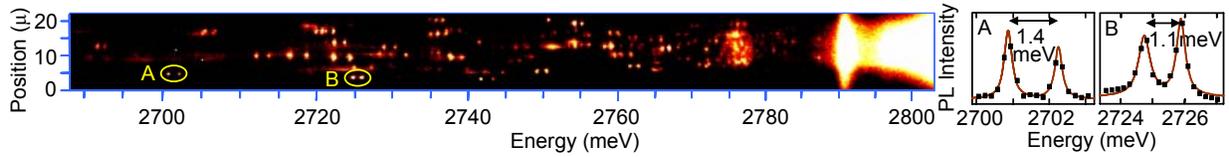

FIG. 1. PL image (laser focus ~20 microns). The vertical axis corresponds to the direction not dispersed by the spectrometer. The two graphs at the right are PL spectra of two specific doublets, labeled A and B in the spectral image.

In Fig. 1, a typical (unpolarized) PL spectral image from our sample is shown. The strong band-edge emission around 2.8 eV is characteristic for high-purity ZnSe. Additionally, many sharp lines appear in the range from 2.6-2.8 eV and are not seen in undoped ZnSe [12]. Such features are the hallmark of excitons confined in all three dimensions and reflect the delta function-like density of states of such a system. Their real linewidth may be much narrower and is limited here by the spectrometer resolution (~400 μeV). Most peaks actually consist of doublets with an average splitting close to 0.6 meV. The PL spectrum of two such doublets is shown in the spectra of Fig. 1.

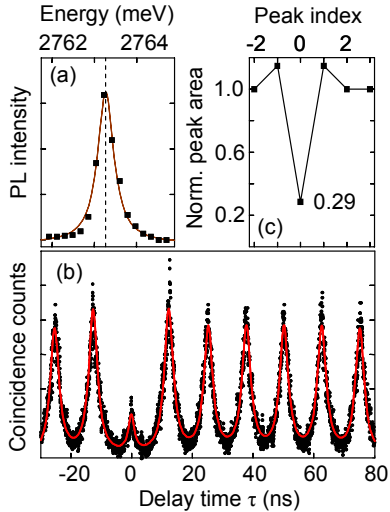

FIG. 2. Photon anti-bunching. (a) Spectrum of the peak whose PL was directed to the HBT setup. (b) Histogram of HBT photon coincidence counts. (c) Normalized peak area extracted from (b).

As has been pointed out in the past [5], observation of sharp PL peaks does not ensure single emitter properties. To verify that the peaks indeed originate from individual quantum emitters, we measured their photon statistics with the HBT setup described above. A typical trace from such a measurement, performed on the peak in Fig. 2(a), is shown in Fig. 2(b). For a light source with Poissonian statistics such as a laser, peaks separated by the laser repetition period would be observed [3]. Here the central peak is suppressed compared to all other peaks [Fig. 2(c)]. The probability of emitting two photons simultaneously is low, which is an intrinsic property of two-level quantum systems.

We next show that the PL doublets of Fig. 1 arise from excitons bound to nearest-neighbor Te pairs. Kuskovsky *et al.* [9] had already argued that single Te atoms could not account for the ~100 meV confinement potential observed in the PL spectra. If excitons bound to *single* Te atoms would in fact give rise to the PL doublets, we would expect a ~500 times greater number of them (we observe 1-5 per μm$^2$) and it would be impossible to isolate them without patterning the sample with submicron apertures or mesas. On the other hand, stochastic estimation of nearest-neighbor Te pairs yields an average concentration of 4 μm$^{-2}$ at our nominal doping, consistent with the experimental observation. Furthermore, the peaks within a doublet are strongly polarized. As shown in Fig. 3(a) for a particular doublet, the emission from the high (low) energy peak is linearly polarized at 90º (0º) in the sample plane, corresponding to the [110] and [$\bar{1}$10] crystallographic axes. In fact, almost all doublets follow identical behavior [Fig. 3(b)]. This is consistent with emission from *nearest-neighbor* pairs of the anion sublattice where Te substitutes Se; the orientation of neighboring Te atoms must be either along [110] or along [$\bar{1}$10], and to each Te$_2$ corresponds one spectral doublet.

Ab-initio density functional calculations confirm the picture given above. Using the one-electron band structure, we derived the relative binding for the electronic states involved in the exciton formation for a single Te and a Te$_2$. Our pseudopotential calculations (local density approximation) [13] were performed in a 64 atoms supercell using 2x2x2 **k**-space sampling and an energy cut-off of 25 Ry. The results give a binding energy ($E_b$) of about 3 meV for the single Te impurity whereas a substantial $E_b$ = 40 meV for the Te$_2$. Even if our calculations are limited by the supercell size they indicate that Te pairs produce a confinement which is one order of magnitude larger than for a single Te. The discrepancy with the observed binding energy for the excitons is likely due to Coulomb and spin-orbit effects.

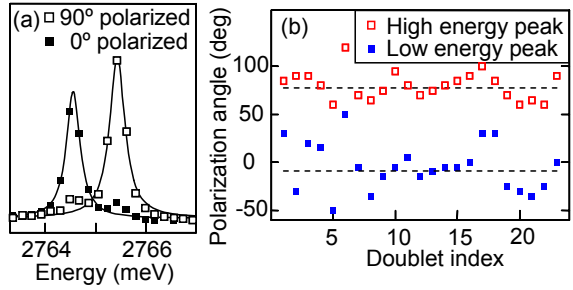

FIG. 3. (a) Polarized spectrum of a single PL doublet. (b) Polarization angle (polarization at PL maximum) for 23 different doublets.

The non-equivalence between [110] and [$\bar{1}$10] is common in MBE-grown (001)-oriented III-V or II-VI compounds where the anion dimerizes along [$\bar{1}$10] on the surface during growth [14]. This explains why the high energy peak is preferentially polarized at 90º [Fig. 3(b)]. We also find no correlation between the emission energy and the magnitude of the splitting (performed on 23 doublets) which is expected if non-nearest neighbor pairs would bind an exciton. Therefore, the inhomogeneity in emission energy and splitting most likely arises from the random strain field created by the Te atoms in the vicinity of each nearest-neighbor Te pair. This is supported by the ab-initio calculation which indicates that the energy gap for ZnSe:Te$_2$ is very sensitive to small changes of the nearest Te-Te bonding length in the supercell. Moreover, the simulation shows that a single Te atom produces a strain field that affects the bonding lengths up to 6%. This interpretation is also consistent with the fact that observed inhomogeneous broadening is larger compared, for example, to the case of GaP:N [15], where the lighter N provides weaker strain.

The D$_{2d}$ quantum-well-like symmetry introduced by the delta-doping allows us to neglect the light holes and consider only the lowest heavy-hole and conduction bands both having $\Gamma_6$ symmetry. The symmetry for the ground state heavy hole exciton states is therefore obtained as $\Gamma_6 \otimes \Gamma_6 = \Gamma_1 \oplus \Gamma_2 \oplus \Gamma_5$. $\Gamma_1$ and $\Gamma_2$ are both 1-dimensional and optically inactive while $\Gamma_5$ is two-dimensional and transforms like the in-plane x, y components of the dipole operator. The short-range e-h exchange interaction splits the dark states $\Gamma_1 \oplus \Gamma_2$ from the optically active $\Gamma_5$. When the symmetry is further reduced to C$_{2v}$ by the presence of Te$_2$, the 2-dimensional representation $\Gamma_5$ becomes $\Gamma_2 \oplus \Gamma_4$ which are split by the long-range part of the e-h exchange. These are the doublets observed in Fig. 1, polarized along the Te$_2$ axis ($\Gamma_2$) and orthogonally to it ($\Gamma_4$).

The optically dark states, though not visible in the PL spectrum, can nevertheless affect the population of the bright states indirectly via a spin-flip process, resulting in a thermal mixture of states. Consequently, the PL decay becomes biexponential [16]. Figure 4(a) shows the time-resolved PL of the high energy peak in the doublet of Fig. 3(a), for different temperatures. While the fast component $\tau_S$ is close to the resolution limit of our setup, the slow component $\tau_L$ is easily resolved and varies significantly with temperature. Such a situation was discussed explicitly for single CdSe colloidal SQDs [16, 17]. A detailed model was developed in Ref. [16] using rate equations within a three-level system, consisting of the bright state $|A\rangle$, the dark state $|F\rangle$ and the ground state $|0\rangle$ (no exciton). Assuming the $|A\rangle \rightarrow |F\rangle$ spin flip rate $\gamma_0$ is larger than the radiative rates $\gamma_A$ and $\gamma_F$, the slow component $\tau_L$ in the biexponential $|A\rangle \rightarrow |0\rangle$ decay reads:

$$\tau_L^{-1} = \frac{\gamma_A + \gamma_F}{2} - \frac{\gamma_A - \gamma_F}{2} \tanh(\frac{\Delta E}{2k_B T})$$

where $T$ is the temperature [inset of Fig. 4(b)]. Note that the result does not depend on the $|A\rangle \rightarrow |F\rangle$ spin flip rate $\gamma_0$ or on the initial conditions. The activation energy $\Delta E$ equals the energy difference between $|A\rangle$ and $|F\rangle$. In Fig. 4(b), the slow decay rate, $\tau_L^{-1}$ extracted from the traces in Fig. 4(a) is plotted versus temperature (filled circles). A fit to this data with the equation above yields $\Delta E$ = 2.4 meV, as well as the intrinsic lifetimes $\gamma_F^{-1}$ = 3.5 ns for the dark state and $\gamma_A^{-1}$ = 0.5 ns for the bright state. The same procedure was applied to the time-resolved data of the low energy peak (same doublet). For this particular Te$_2$ we thus obtain the energy diagram shown in Fig. 4(c).

We also probed other Te$_2$ and found comparable lifetimes, typically $\gamma_F^{-1}$ ~ 2-4 ns and $\gamma_A^{-1}$ < 0.5 ns. The data represented by x's in Fig. 4(b), for example, was recorded on the high energy peak of a different doublet. Note that the data closely follows the hyperbolic tangent function. Unlike colloidal SQDs [16], single Te$_2$ exhibit a systematic bi-exponential PL decay. This material system is thus unique in that both the bright states, and the effect of the dark states can be observed without an external magnetic field. This picture could nevertheless easily be complemented and confirmed by magneto-optical measurements.

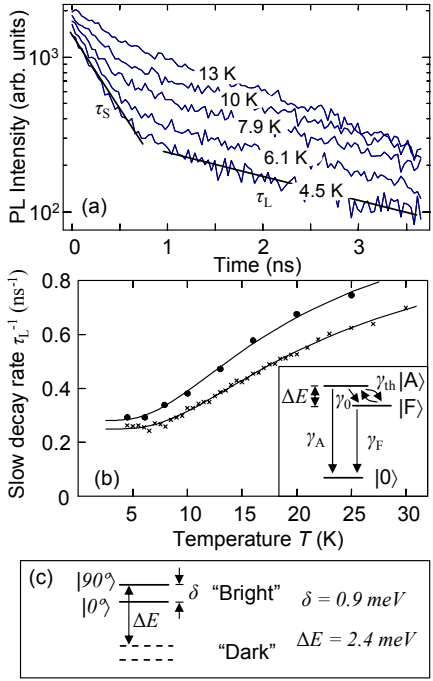

FIG. 4. (a) Time-resolved traces, for various temperatures, of the PL from the high energy peak in Fig. 3(a). (b) Slow decay rate $\tau_L^{-1}$ versus temperature for the data in (a) (filled circles) and for data from a completely different doublet (x's). The inset depicts the Te$_2$ energy level diagram. State $|0\rangle$ represents the crystal ground-state (no exciton), $\gamma_0$ is the zero temperature $|A\rangle \rightarrow |F\rangle$ relaxation rate, and $\gamma_{th} = \gamma_0 N_B$. $N_B = \frac{1}{e^{\Delta E/k_B T} - 1}$ is the Bose-Einstein phonon number at temperature $T$. (c) Summary of the fine-structure of one particular Te$_2$ probed.

In summary, using PL spectroscopy we have investigated the zero-dimensional semiconductor system consisting of Te impurities in bulk ZnSe, in the limit of very low Te concentration. We unambiguously find that in such a sample, PL doublets originate from the recombination of excitons bound to nearest-neighbor Te pairs whose fine-structure is manifested in the spectral and temporal PL. Such a system turns out to be very promising for studies of zero-dimensional excitons, since it possesses possible advantages over SQD systems, such as a well-defined structure and symmetry, and should be easier to model theoretically. It could also be suited for quantum optical experiments where large transition dipole moments are essential [5], if proven to provide sufficiently long coherence times.

The authors thank Prof. H. Mariette and Prof. J. W. Keto for fruitful discussions and Prof. L. M. Smith for advice regarding the solid immersion lens. This work was supported by NSF (DMR-0210383, DMR-0306239 and DMR-0312491), Texas Advanced Technology program, and the W.M. Keck Foundation.

*Corresponding Author. E-mail: shih@physics.utexas.edu